\documentclass[10pt,twocolumn,letterpaper]{article}

\usepackage[pagenumbers]{cvpr} 
\usepackage[accsupp]{axessibility}
\definecolor{cvprblue}{rgb}{0.21,0.49,0.74}
\usepackage[pagebackref,breaklinks,colorlinks,allcolors=cvprblue]{hyperref}
\usepackage{xurl}
\usepackage{listings}
\usepackage{xcolor}
\usepackage{tcolorbox}
\tcbuselibrary{skins}
\usepackage{listings}
\usepackage{tikz}
\def\paperID{19} 
\def\confName{CVPR}
\def\confYear{2026}

\title{Authenticated Contradictions from Desynchronized Provenance and Watermarking}

\author{
  Alexander Nemecek$^{1}$ \quad
  Hengzhi He$^{2}$ \quad
  Guang Cheng$^{2}$ \quad
  Erman Ayday$^{1}$\thanks{Correspondence: exa208@case.edu.} \\[4pt]
  $^{1}$\,Case Western Reserve University \quad
  $^{2}$\,University of California, Los Angeles
}

\begin{document}
\maketitle

\begin{abstract}
Cryptographic provenance standards such as C2PA and invisible watermarking are positioned as complementary defenses for content authentication, yet the two verification layers are technically independent: neither conditions on the output of the other. This work formalizes and empirically demonstrates the \textit{Integrity Clash}, a condition in which a digital asset carries a cryptographically valid C2PA manifest asserting human authorship while its pixels simultaneously carry a watermark identifying it as AI-generated, with both signals passing their respective verification checks in isolation. We construct metadata washing workflows that produce these authenticated fakes through standard editing pipelines, requiring no cryptographic compromise, only the semantic omission of a single assertion field permitted by the current C2PA specification. To close this gap, we propose a cross-layer audit protocol that jointly evaluates provenance metadata and watermark detection status, achieving 100\% classification accuracy across 3,500 test images spanning four conflict-matrix states and three realistic perturbation conditions. Our results demonstrate that the gap between these verification layers is unnecessary and technically straightforward to close. Our code is available at \url{https://github.com/ANCP2021/integrity-clash}.
\end{abstract}

\section{Introduction}
\label{sec:intro}
Generative artificial intelligence (AI) systems now produce synthetic media that is perceptually indistinguishable from authentic content across text, audio, and image modalities~\cite{nightingale2022ai, cooke2025good}. This capability introduces pressing security concerns as synthetic media can be weaponized for misinformation campaigns, used to infringe on intellectual property, and exploited for identity fraud~\cite{ferrara2024genai, monteith2024artificial, CDRF2025synthetic}. In response, particularly within the image domain, two parallel verification paradigms have emerged. The first, standards-based cryptographic provenance exemplified by the Coalition for Content Provenance and Authenticity (C2PA)~\cite{C2PA2026spec} and the Content Authenticity Initiative (CAI)~\cite{CAI2026website}, attaches cryptographically signed metadata manifests to digital assets, declaring their creation and editing history. The second, invisible watermarking deployed through systems such as Google's SynthID~\cite{GoogleDeepMind2026synthid} and Meta's Meta Seal~\cite{MetaSeal2026}, embeds imperceptible origin signals directly into the pixel data. Industry roadmaps and technical guidance from both the C2PA specification and major platform deployers position these two systems as complementary, with their combination framed as a defense-in-depth strategy for content authentication~\cite{C2PA2025ux, Guinard2024c2pa}.

This framing, however, obscures a structural vulnerability that follows directly from the architecture of both systems. The two verification layers are technically independent where C2PA manifest validation is a cryptographic signature check over metadata~\cite{c2pa2024explainer, c2pa2023security, simmons2024interoperable}, while watermark detection is a signal-decoding operation over pixel data~\cite{zhong2023brief, zhu2018hidden}. Neither procedure conditions on the output of the other. As a result, a digital asset can carry a cryptographically valid C2PA manifest asserting one provenance history while its pixels simultaneously carry a watermark signal that contradicts that history, with both signals passing their respective verification checks in isolation. In this paper, we formalize and empirically demonstrate this condition, which we refer to as the \textit{Integrity Clash}. Concretely, an AI-generated image can be passed through a C2PA-compliant editing tool, signed with valid credentials that assert only a human editing action, and emerge with a valid provenance manifest while still containing a watermark that identifies it as synthetically generated. A verifier that inspects only one layer receives an incomplete account of the asset's origin.

Two factors make this vulnerability practically relevant. First, adoption of both systems is accelerating. C2PA manifests are supported in Adobe Photoshop~\cite{adobe2024photoshop_cc}, Nikon~\cite{nikon2024z6iii_c2pa}, Leica~\cite{leica2023m11p}, and Sony camera firmware~\cite{sony2024c2pa}; invisible watermarking is deployed at platform scale by Google~\cite{gowal2025synthid} and Meta~\cite{fernandez2023stable}, among others. The number of assets that carry both verification signals is therefore increasing. Second, to our knowledge, no deployed verification workflow jointly evaluates both layers for semantic consistency. Platforms that display C2PA-based provenance indicators do not cross-reference watermark detection outputs~\cite{parsons2024durable, c2pa2024softbinding, fairoze2025difficulty}, and watermark detectors operate without access to the asset's declared provenance~\cite{srinivasan2024detecting, kassis2025unmarker}. The \textit{Integrity Clash} is therefore not a theoretical concern but a testable condition in currently deployed infrastructure, and the experiments in this paper are designed to instantiate and detect it.

Prior work has examined each verification layer in isolation. Research on watermarking evaluates embedding robustness under compression, cropping, and adversarial attack~\cite{an2024waves, zhao2024invisible, saberi2023robustness}, while research on provenance standards analyzes manifest integrity, soft-binding collision risks, and metadata stripping vulnerabilities~\cite{c2pa2023security, c2pa2024softbinding}. Recent empirical audits have documented fragmented adoption of these systems across generative AI platforms~\cite{nemecek2025watermarking, rijsbosch2025adoption}, and preliminary frameworks have proposed multi-signal consistency checks as a forensic methodology~\cite{simmons2024interoperable, itu2024deepfakes}. However, no existing work empirically demonstrates the cross-layer contradiction as a concrete, reproducible vulnerability in deployed pipelines, nor provides a systematic protocol for detecting such contradictions in deployed pipelines. This paper addresses that gap. We make three contributions:
\begin{itemize}
    \item We \textbf{formalize the Integrity Clash} through a threat model and a conflict matrix that categorizes all combinations of provenance and watermark signals into four states, identifying the \textit{Authenticated Fake} quadrant as the inconsistency.
    
    \item We \textbf{construct metadata washing workflows} using openly available tools, showing that authenticated fakes can be produced through standard editing pipelines with no cryptographic compromise.
    
    \item We \textbf{propose and evaluate a cross-layer audit protocol} that jointly verifies C2PA metadata and watermark signals, detecting contradictions that single-layer verification cannot identify by design.
\end{itemize}
Taken together, these contributions demonstrate that provenance metadata and pixel-level watermarking, as currently deployed, can produce semantically contradictory verification outcomes on the same asset, and that a joint audit protocol can reliably surface these contradictions.

\section{Background and Related Work}
\label{sec:background}
This section establishes the technical properties of the two verification layers central to our analysis and positions our work against prior literature.

\subsection{Standards-Based Provenance (C2PA)}
\label{subsec:c2pa}
The C2PA standard encodes provenance as a \textit{manifest}: a digitally signed block of metadata containing assertions about an asset's origin and editing history~\cite{cai2024manifests}. To bind a manifest to specific pixel data, the specification defines \textit{hard bindings} (cryptographic hashes over the asset's bytes) and \textit{soft bindings} (perceptual hashes or embedded watermarks)~\cite{C2PA2026spec}. Any re-encoding, recompression, or format conversion invalidates the hash even when the image is visually unchanged, pushing real-world workflows toward soft bindings, which the specification acknowledges are vulnerable to collision-based attacks~\cite{c2pa2023security}. For this work, the architectural property is that C2PA is a \textit{declarative} system where a valid signature certifies that the metadata has not been modified since signing and can be attributed to a specific key, but it does not certify the semantic truth of the assertions. A manifest claiming human authorship may be cryptographically valid regardless of whether the underlying pixels were generated by an AI model.

Recent security analyses have further highlighted that C2PA manifests remain vulnerable to re-signing attacks, where adversaries strip and reattach manifests with altered assertions, and to soft-binding collisions that allow adversarially crafted content to inherit legitimate provenance chains~\cite{krawetz2024c2pa}. Joint guidance from national cybersecurity agencies acknowledges that while C2PA strengthens attribution, its effectiveness depends on ecosystem-wide adoption and does not prevent semantically misleading assertions from being signed by legitimate actors~\cite{nsa2025contentcredentials}. These findings reinforce the property that cryptographic validity does not entail semantic truthfulness, and authenticated yet misleading provenance claims can arise when C2PA is used in isolation.

\subsection{Invisible Watermarking for Generative AI}
\label{subsec:watermarking}
Invisible watermarking embeds imperceptible origin signals directly into an image's pixel data. Current approaches follow two paradigms, \textit{latent-space} methods that embed the watermark into the latent representation during the diffusion sampling process (e.g., Tree-Ring~\cite{wen2023tree}, GaussianShading~\cite{yang2024gaussian}), and \textit{post-hoc} methods that embed payload bits as a post-processing step, either via a learned encoder network (e.g., SynthID-Image~\cite{gowal2025synthid}, StegaStamp~\cite{tancik2020stegastamp}, TrustMark~\cite{bui2025trustmark}) or by modifying selected coefficients in a transform domain~\cite{al2007combined, gunjal2011secured}. Our experiments use a post-hoc method, Pixel Seal~\cite{souvcek2025pixel}, part of Meta's Meta Seal suite~\cite{MetaSeal2026}, which represents the current state of the art for robustness and imperceptibility in the image domain. The key property for this work is that watermarks are bound to pixel data, not to file metadata. Once embedded, the signal persists through recompression, format conversion, and metadata stripping, degrading only when processing substantially disrupts the encoded frequency bands~\cite{liang2024robust}.

Recent work has shown that invisible watermarks are not immutable. Benchmarks such as WAVES~\cite{an2024waves} evaluate robustness under compression, rescaling, and noise, and adversarial studies have demonstrated practical removal and transplantation attacks against open-source watermarking tools~\cite{lukas2023leveraging, nam2020attack}. The winning system in the NeurIPS 2024 Invisible Watermark Removal Challenge~\cite{shamshad2025first, ding2024erasing}, for example, uses generative-model-based regeneration to strip a variety of watermark signals while maintaining high perceptual quality, and Zhao et al.~\cite{zhao2024invisible} prove that a broad class of invisible image watermarks are removable using generative AI. These findings establish that watermark signals can be weakened under determined attackers, but also confirm that under standard editorial workflows, watermarks remain persistent to produce detectable cross-layer inconsistencies with co-existing metadata.

\subsection{Related Work}
\label{subsec:related}
The robustness and vulnerability properties reviewed have been studied, but almost exclusively within a single verification layer. On the provenance side, the C2PA security considerations documents~\cite{c2pa2023security} enumerate threats including manifest stripping, re-signing, and soft-binding collisions, while others argue that C2PA's reliance on voluntary adoption renders it insufficient as a standalone defense against misinformation~\cite{locker2025overpromising}. Hardware-anchored approaches propose binding provenance to the capture device, however these works treat the watermark layer as out of scope~\cite{jang2025signing}.

Ecosystem audits have begun to document the gap between these layers. Rijsbosch et al.~\cite{rijsbosch2025adoption} report that only 38\% of surveyed AI image generators implement adequate watermarking and just 18\% integrate with C2PA-style labeling standards, establishing that the fragmentation enabling cross-layer inconsistencies is empirically widespread. Nemecek et al.~\cite{nemecek2025watermarking} argue that without independent verification across technical, audit, and enforcement layers, watermarking serves as symbolic compliance rather than effective oversight. Simmons and Winograd~\cite{simmons2024interoperable} define success and failure scenarios when provenance and watermark signals co-exist in broadcast media, while agent-based forensic pipelines have been explored for multimodal detection~\cite{liang2025evidence}. These contributions identify the fragmentation problem or propose frameworks, but none demonstrate the cross-layer contradiction as a reproducible specification-level vulnerability or evaluate a detection protocol against it.

Industry and standards-focused analyses reinforce this picture at a systems level. Recent work on media integrity~\cite{young2026media} concludes that high-confidence validation requires aligning multiple signals rather than trusting any single mechanism in isolation, and the Origin Lens framework~\cite{loth2026origin} advocates combined provenance verification and AI detection but does not examine cases where a valid manifest coexists with a contradictory watermark. None, to our knowledge, empirically demonstrates the cross-layer contradiction as a concrete, reproducible vulnerability in deployed tools, nor provides a systematic protocol for detecting such contradictions. Our work addresses this gap by constructing workflows that exploit this specification-level omission to produce authenticated fakes, and evaluating a cross-layer audit protocol that jointly reasons over provenance metadata and pixel-level watermark signals.

\section{Threat Model}
\label{sec:threat_model}
We model a digital image as carrying two independent verification signals. The \textit{metadata layer} $M$ is the content of a C2PA manifest, including its validation status and semantic assertions (e.g., origin, editing actions, software agent). The \textit{watermark layer} $W$ is the output of a watermark detector applied to the image's pixel data, returning either a detection of the embedded payload (indicating synthetic origin) or no detection. Each layer has its own verification procedure: $M$ is validated through cryptographic signature checks, and $W$ through signal decoding. These procedures are independent as the validity of $M$ is never conditioned on the state of $W$, nor vice versa.

We define four states over the pair $(M, W)$, which we systematize as a conflict matrix in Section~\ref{subsec:conflict_matrix}. The two states central to this work are the \textit{synchronized state}, where both layers are present and semantically consistent, and the \textit{synchronization gap}, where $M$ passes cryptographic validation and asserts a provenance history that does not disclose AI generation, while $W$ detects an AI-origin watermark. An asset in this latter state, cryptographically valid yet semantically contradictory, is an \textit{authenticated fake}.

The adversary we consider can generate AI images using publicly available models, embed or preserve watermarks using open-source tools, and pass images through C2PA-compliant signing workflows to attach valid manifests. The adversary \textit{cannot} forge cryptographic signatures, compromise certificate authorities, or break standard cryptographic primitives. The resulting vulnerability is therefore not cryptanalytic but procedural, as the adversary simply omits the AI-origin assertion from the manifest, which is permitted by the specification since C2PA does not mandate that signers declare generative origins~\cite{C2PA2026spec}. While the \textit{Integrity Clash} can arise through other paths such as soft-binding collisions, this work focuses on \textit{provenance laundering}, in which watermarked AI-generated content is passed through a legitimate signing workflow without disclosure of its synthetic origin, as the most operationally straightforward vector.

\subsection{Cross-Layer Consistency States}
\label{subsec:conflict_matrix}
To systematize the possible relationships between $M$ and $W$, we define a conflict matrix over two binary dimensions: whether a valid C2PA manifest is present, and whether an AI-origin watermark is detected. This yields four quadrants, illustrated in Figure~\ref{fig:conflict_matrix}:
\begin{itemize}
    \item \textbf{Q1 Silent Zone:} Neither a valid manifest nor a detectable watermark is present. The asset carries no machine-readable authenticity signal from either layer.
    \item \textbf{Q2 Fragile Provenance:} $M$ is absent or invalid; $W$ detects an AI watermark. The watermark provides evidence of synthetic origin, but no metadata context exists to contradict or confirm it.
    \item \textbf{Q3 Authenticated Content:} $M$ is valid; $W$ does not detect a watermark. The manifest is the sole verification signal, consistent with a standard human-authored asset carrying a valid provenance credential.
    \item \textbf{Q4 Dual Signal:} $M$ is valid and $W$ detects an AI watermark. Both verification layers are present on the same asset, and their semantic relationship determines whether the provenance claim is consistent or contradictory.
\end{itemize}

\begin{figure}[t]
\centering
\begin{tikzpicture}[scale=0.92, every node/.style={scale=0.92},
  cell/.style={minimum width=3.8cm, minimum height=1.6cm, align=center, font=\small},
  header/.style={minimum width=3.8cm, minimum height=0.7cm, align=center, font=\small},
  rowlabel/.style={minimum width=0cm, minimum height=1.6cm, align=center, font=\small}
]
\node[header] at (2.9, 2.2) {No Watermark};
\node[header] at (6.7, 2.2) {Watermark Detected};
\node[rowlabel] at (0.2, 1.0) {No valid \\ C2PA};
\node[rowlabel] at (0.2, -0.6) {Valid \\ C2PA};
\node[cell, fill=gray!15, draw=black, line width=0.5pt] at (2.9, 1.0) {
  \textbf{Q1}\\Silent Zone
};
\node[cell, fill=yellow!20, draw=black, line width=0.5pt] at (6.7, 1.0) {
  \textbf{Q2}\\Fragile Provenance
};
\node[cell, fill=blue!8, draw=black, line width=0.5pt] at (2.9, -0.6) {
  \textbf{Q3}\\Authenticated\\Content
};
\begin{scope}
  \clip (4.8,-1.4) rectangle (8.6,0.2);
  \fill[green!20] (4.8,0.2) -- (8.6,0.2) -- (4.8,-1.4) -- cycle;
  \fill[red!18] (8.6,0.2) -- (8.6,-1.4) -- (4.8,-1.4) -- cycle;
\end{scope}
\draw[black, line width=0.5pt] (4.8,0.2) rectangle (8.6,-1.4);
\node[font=\small, align=center] at (5.5,-0.45) {
  \textbf{Q4a}\\[-1pt]{\scriptsize Verified}\\[-2pt]{\scriptsize Synthetic}
};
\node[font=\small, align=center] at (7.8,-0.75) {
  \textbf{Q4b}\\[-1pt]{\scriptsize Authenticated}\\[-2pt]{\scriptsize Fake}
};
\end{tikzpicture}
\caption{Cross-layer conflict matrix. Q4 splits into Q4a (Verified Synthetic), where the manifest discloses AI generation, and Q4b (Authenticated Fake), where it does not. The transition requires no cryptographic compromise, only semantic omission of the AI origin assertion.}
\label{fig:conflict_matrix}
\end{figure}

Q4 is the quadrant central to this work. When the manifest discloses AI generation and the watermark confirms it, we refer to this state as \textbf{Q4a (Verified Synthetic)}, the intended behavior of a fully functioning authenticity pipeline. When the manifest omits AI disclosure while the watermark detects synthetic origin, we refer to this state as \textbf{Q4b (Authenticated Fake)}, the \textit{Integrity Clash}. We note an adversary could alternatively strip the watermark to reach Q3 or remove the manifest to reach Q2, but both strategies leave an observable absence in one verification layer. The Q4b state is challenging to detect precisely because both signals remain intact and pass their respective checks in isolation.

\section{Method}
\label{sec:method}
To instantiate and detect the \textit{Integrity Clash}, we construct four experimental pipelines that systematically populate each quadrant of the conflict matrix. Each pipeline applies a distinct combination of watermarking and C2PA signing to the same base image set, enabling controlled comparison across verification states. The following subsections describe the dataset, watermarking procedure, signing workflows, robustness perturbations, and audit protocol.

\subsection{Dataset}\label{sec:dataset}
We generate 500 synthetic images using \texttt{Stable-Diffusion-XL-Base-1.0} (SDXL)~\cite{podell2023sdxl} at a resolution of 1024$\times$1024 pixels in PNG format. Images are sampled from the Parti-Prompts benchmark~\cite{yu2022scaling}, which provides a structured set of text prompts spanning diverse semantic categories and compositional challenge dimensions. To ensure coverage without overrepresentation of any single category or challenge type, prompts are selected via a round-robin allocation that distributes samples approximately equally across categories before cycling over challenge aspects within each. The built-in SDXL watermarker is explicitly disabled to ensure that any watermark signal present in downstream pipelines is attributable exclusively to our controlled embedding step. All prompts and random seeds are recorded to support reproducibility.

\subsection{Watermarking}\label{sec:watermarking}
All 500 images are embedded with an invisible watermark using Pixel Seal~\cite{souvcek2025pixel}, part of Meta's Meta Seal~\cite{MetaSeal2026} watermarking suite and its highest-performing model for robustness and imperceptibility in the image domain. Pixel Seal is a post-hoc learned encoder-decoder framework that embeds a fixed 256-bit payload into each image as a post-processing step without modifying the underlying generative model.

Watermark detection is performed by decoding the 256-bit payload from each image and computing bit accuracy as the proportion of bits correctly recovered relative to the embedded target. Detection decisions are made by thresholding bit accuracy at 0.75, the midpoint between chance-level performance (approximately 0.50 for a random 256-bit signal) and perfect recovery (1.0). This threshold is chosen to remain fair under degradation of a watermark that survives realistic perturbations with bit accuracy above 0.75 is considered detected, while one degraded below this boundary is treated as absent.

\subsection{C2PA Signing}\label{sec:C2PA}
C2PA manifests are attached to images using the official \texttt{c2pa-python} bindings from the Content Authenticity Initiative\footnote{\url{https://github.com/contentauth/c2pa-python}}. All signing operations use a self-signed ECDSA P-256 certificate chain consisting of a root CA certificate and a leaf signing certificate, generated with OpenSSL and combined into a chain with the signing certificate first. A DigiCert timestamp authority (\url{http://timestamp.digicert.com}) is included at signing time to bind each manifest to a trusted timestamp. The use of a self-signed CA is deliberate and consistent with the threat model. The vulnerability under investigation is semantic omission, not certificate forgery, and a self-signed chain produces a cryptographically valid manifest that exercises the same verification pathway as a commercially trusted credential.

Two manifest templates are used across pipelines, differing only in their declared action and software agent fields. The AI disclosure manifest declares a \texttt{c2pa.created} action, specifies \texttt{trainedAlgorithmicMedia} as the \texttt{digitalSourceType}, and identifies \texttt{StableDiffusionXL/1.0} as the software agent, constituting a complete and honest declaration of synthetic origin. The human-edited manifest declares a \texttt{c2pa.edited} action, identifies \texttt{PhotoEditor/2.0} as the software agent, and omits the \texttt{digitalSourceType} field entirely, making no disclosure of AI involvement. Both manifests are signed with the same certificate, key, and timestamp authority. Figure~\ref{fig:manifest_diff} presents the two templates as a structured diff. The complete difference between an honestly declared AI-generated image and an authenticated fake reduces to the omission of a single assertion field, an omission that is permitted by the C2PA specification, which does not mandate disclosure of generative origins~\cite{C2PA2026spec}.

\begin{table*}[t]
\centering
\small
\begin{tabular}{llll}
\toprule
\textbf{Pipeline} & \textbf{Watermark} & \textbf{C2PA Manifest} & \textbf{Quadrant} \\
\midrule
Baseline                          & None      & None                                          & Q1 \\
Watermarked                       & Pixel Seal & None                                          & Q2 \\
Honest Manifest                   & Pixel Seal & \texttt{c2pa.created}, AI disclosed            & Q4a \\
Misleading Manifest               & Pixel Seal & \texttt{c2pa.edited}, no AI disclosure          & Q4b \\
\midrule
Misleading + JPEG Q80              & Pixel Seal & \texttt{c2pa.edited}, no AI disclosure & Q4b* \\
Misleading + Crop 10\% + resize    & Pixel Seal & \texttt{c2pa.edited}, no AI disclosure & Q4b* \\
Misleading + Screenshot simulation & Pixel Seal & \texttt{c2pa.edited}, no AI disclosure & Q4b* \\
\bottomrule
\multicolumn{4}{l}{\footnotesize *Expected Q4b if watermark survives perturbation; degrades to Q3 if bit accuracy falls below 0.75.}
\end{tabular}
\caption{Each pipeline applies a distinct combination of watermarking and C2PA signing to the same 500-image base set, mapping to one quadrant of the conflict matrix. The Honest and Misleading pipelines take identical watermarked images as input; the only difference is the manifest template used at signing time. Robustness variants apply perturbations before signing with the misleading manifest.}
\label{tab:pipeline_summary}
\end{table*}

\begin{figure}[t]
\centering

\begin{tcolorbox}[
  colback=green!6!white,
  colframe=green!50!black,
  title={\small\bfseries Q1: Honest AI Disclosure},
  coltitle=white,
  attach boxed title to top left={yshift=-2mm, xshift=4mm},
  boxed title style={colback=green!50!black, colframe=green!50!black,
    rounded corners, sharp corners=downhill},
  rounded corners, arc=3pt, boxrule=0.6pt,
  top=4pt, bottom=4pt, left=6pt, right=6pt
]\footnotesize
\begin{verbatim}
{
  "action": "c2pa.created",
  "digitalSourceType":
    "...trainedAlgorithmicMedia",
  "softwareAgent":
    "StableDiffusionXL/1.0"
}
\end{verbatim}
\end{tcolorbox}
\begin{tcolorbox}[
  colback=red!5!white,
  colframe=red!60!black,
  title={\small\bfseries Q2: Authenticated Fake},
  coltitle=white,
  attach boxed title to top left={yshift=-2mm, xshift=4mm},
  boxed title style={colback=red!60!black, colframe=red!60!black,
    rounded corners, sharp corners=downhill},
  rounded corners, arc=3pt, boxrule=0.6pt,
  top=4pt, bottom=4pt, left=6pt, right=6pt
]\footnotesize
\begin{verbatim}
{
  "action": "c2pa.edited",
  "softwareAgent": "PhotoEditor/2.0"
  // digitalSourceType: ABSENT
}
\end{verbatim}
\end{tcolorbox}

\caption{Comparison of an honestly declared AI-generated image manifest
(Q4a, top) and an authenticated fake (Q4b, bottom). The \texttt{digitalSourceType}
field is entirely absent in the misleading manifest, and the action and software
agent are replaced with generic editing descriptors. The workflow requires no cryptographic compromise, only the semantic omission of the AI origin assertion, which the C2PA specification does not mandate~\cite{C2PA2026spec}.}
\label{fig:manifest_diff}
\end{figure}

\subsection{Robustness Perturbations}\label{sec:robustness}
To evaluate whether the \textit{Integrity Clash} persists under realistic image processing, we apply three perturbation types to the watermarked images \emph{before} signing with the human-edited manifest. Each perturbation models a common editorial or platform-mediated transformation that an image may undergo between generation and publication. 
We apply (i)~JPEG compression at quality 80, simulating the most common lossy re-encoding on the web, (ii)~a 10\% center crop followed by resizing back to 1024$\times$1024, simulating editorial reframing, and (iii)~a screenshot simulation pipeline consisting of downscaling to 75\% resolution, JPEG compression at quality 70, and upscaling back to the original dimensions, approximating the cumulative degradation of a social-media sharing cycle. Each perturbation is applied independently to all 500 watermarked images, yielding 1,500 additional test cases. Only the human-edited manifest is used for signing, since the robustness question is whether the cross-layer contradiction survives degradation, not whether honest disclosure remains intact.

\subsection{Cross-Layer Audit Protocol}\label{sec:audit}
Because existing verification workflows evaluate provenance metadata and watermark signals independently (Section~\ref{sec:threat_model}), the contradictory state Q4b is invisible to both. The audit protocol proposed here closes this gap by jointly evaluating both layers and applying consistency rules derived from the conflict matrix.

For each image, the protocol extracts two signals: (1)~whether a cryptographically valid C2PA manifest is present and, if so, whether its assertions include an AI-origin disclosure (specifically, the presence of \texttt{trainedAlgorithmicMedia} as \texttt{digitalSourceType}), and (2)~whether the watermark bit accuracy exceeds the detection threshold of 0.75. The image is then classified into one of the conflict-matrix states. If neither a valid manifest nor a detectable watermark is present, the image is classified as Q1 (Silent Zone). If no valid manifest is present but a watermark is detected, the image is classified as Q2 (Fragile Provenance). If a valid manifest is present and no watermark is detected, the image is classified as Q3 (Authenticated Content). If both a valid manifest and a detected watermark are present, the classification depends on manifest content, Q4a (Verified Synthetic) if the manifest discloses AI generation, and Q4b (Authenticated Fake) if it does not. A Q4b classification constitutes the protocol's positive detection of a cross-layer contradiction.

Table~\ref{tab:pipeline_summary} summarizes the four core experimental pipelines and the three robustness variants, mapping each to its expected quadrant. The protocol is applied to all pipeline outputs and evaluated against known ground truth, with Q4b classifications as the positive class for computing detection metrics.

\begin{table*}[t]
\centering
\small
\begin{tabular}{lccccc}
\toprule
Pipeline & $N$ & C2PA Valid (\%) & Avg. Bit Acc. & Min. Bit Acc. & Classified Correctly (\%) \\
\midrule
Baseline        & 500 & 0.0   & 0.502 & 0.410 & 100.0 \\
Watermarked     & 500 & 0.0   & 0.999 & 0.973 & 100.0 \\
Honest Manifest & 500 & 100.0 & 0.999 & 0.973 & 100.0 \\
Misleading Manifest & 500 & 100.0 & 0.999 & 0.973 & 100.0 \\
\bottomrule
\end{tabular}
\caption{Core pipeline results. All pipelines share identical bit accuracy (0.999), confirming that C2PA signing does not alter pixel data. All images are classified into their expected conflict-matrix quadrant.}
\label{tab:core-results}
\end{table*}

\section{Evaluation}
\label{sec:evaluation}
This section reports results across all experimental pipelines defined in Section~\ref{sec:method}. We first present core pipeline outcomes and validate them against deployed verification infrastructure (Section~\ref{sec:core}), then evaluate the cross-layer audit protocol, including watermark persistence under realistic perturbations (Section~\ref{sec:CLAP}).

\subsection{Core Pipeline}\label{sec:core}
Table~\ref{tab:core-results} summarizes verification outcomes across the four core pipelines. The baseline images, which carry no embedded watermark, produce a mean bit accuracy of 0.502, consistent with chance-level decoding of a 256-bit payload, and the minimum observed value of 0.410 remains well below the 0.75 detection threshold, confirming that the detector does not produce false positives on unmarked content. The three watermarked pipelines share identical bit accuracy statistics (mean 0.999, minimum 0.973), confirming that C2PA manifest signing operates exclusively on file-level metadata and does not modify pixel data. Our main result is the ``Misleading Manifest'' pipeline as all 500 images carry a cryptographically valid C2PA manifest asserting only a human editing action while simultaneously carrying a watermark identifying them as synthetically generated. The two manifest pipelines take identical watermarked images as input and differ only in which template is applied at signing time, reducing the distance between a correctly labeled synthetic image and an authenticated fake to the omission of a single assertion field.

\begin{figure}[t]
\centering
\includegraphics[width=\linewidth]{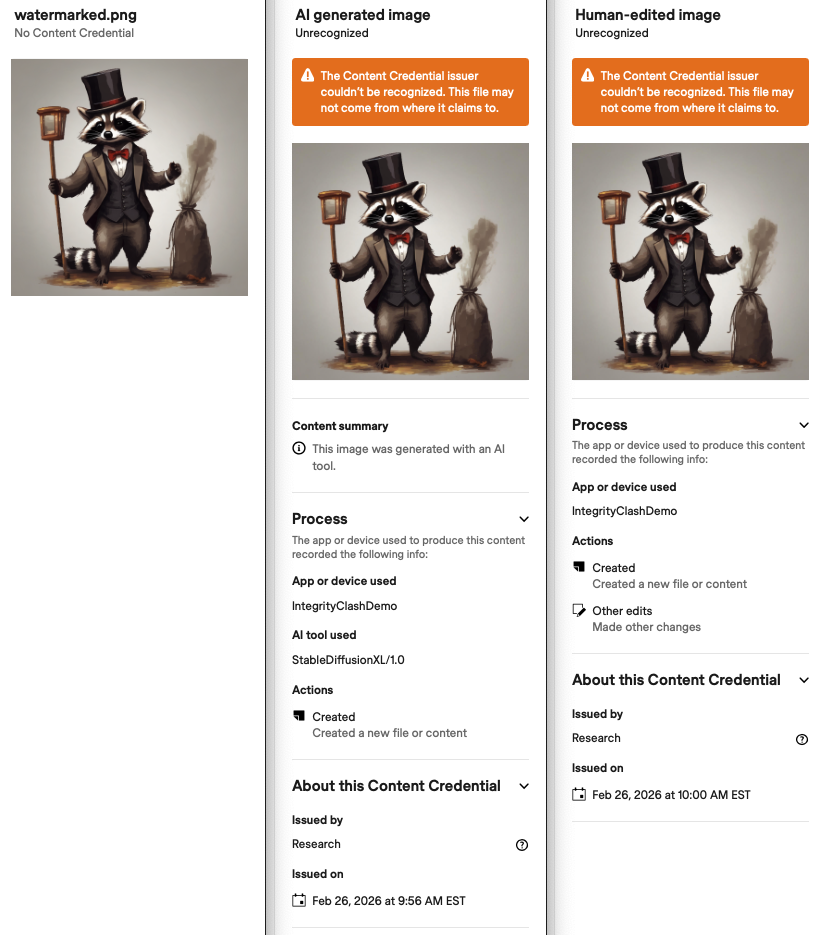}
\caption{The same AI-generated, watermarked image as displayed by the Content Credentials Verify tool under three conditions: the original watermarked image with no manifest attached, reporting no content credential (left), honest AI-disclosure manifest (middle), correctly identifying the image as AI-generated, and misleading human-edited manifest (right), displaying the image as human-edited with no mention of AI involvement. The output is determined entirely by the attached metadata; watermark signals are not inspected. The issuer warning refers only to our research certificate and would not appear with a trusted credential.}
\label{fig:verify-screenshots}
\end{figure}

Figure~\ref{fig:verify-screenshots} illustrates this result against deployed verification infrastructure. We submitted the same AI-generated, watermarked image to the Content Credentials Verify\footnote{\url{https://verify.contentauthenticity.org}} tool under three conditions: with no manifest attached, with the honest AI-disclosure manifest, and with the misleading human-edited manifest. The verifier's output changes entirely based on the attached metadata. When no manifest is present, the tool reports no content credential. When the honest manifest is attached, the tool correctly identifies the image as AI-generated and surfaces the generative model. When the misleading manifest is attached, the tool displays the image as human-edited with no mention of AI involvement. In both signed cases, the only warning raised concerns the certificate issuer, which is unrecognized because our research certificate is not in the platform's trust store. The verifier has no mechanism to inspect or cross-reference the pixel-level watermark signal. A CA-issued certificate from a C2PA-compliant tool would eliminate the issuer warning while the underlying semantic contradiction remains undetected. We note that Q3 (Authenticated Content), in which a valid manifest is present but no watermark is detected, is not instantiated as a separate pipeline. The baseline images confirm the detector's behavior on unwatermarked content, and Q3 would require either a human-captured image signed with C2PA or the removal of the watermark before signing, neither of which is the vulnerability under investigation. 

\subsection{Cross-Layer Audit Protocol}
\label{sec:CLAP}
\begin{figure}[t]
\centering
\includegraphics[width=\linewidth]{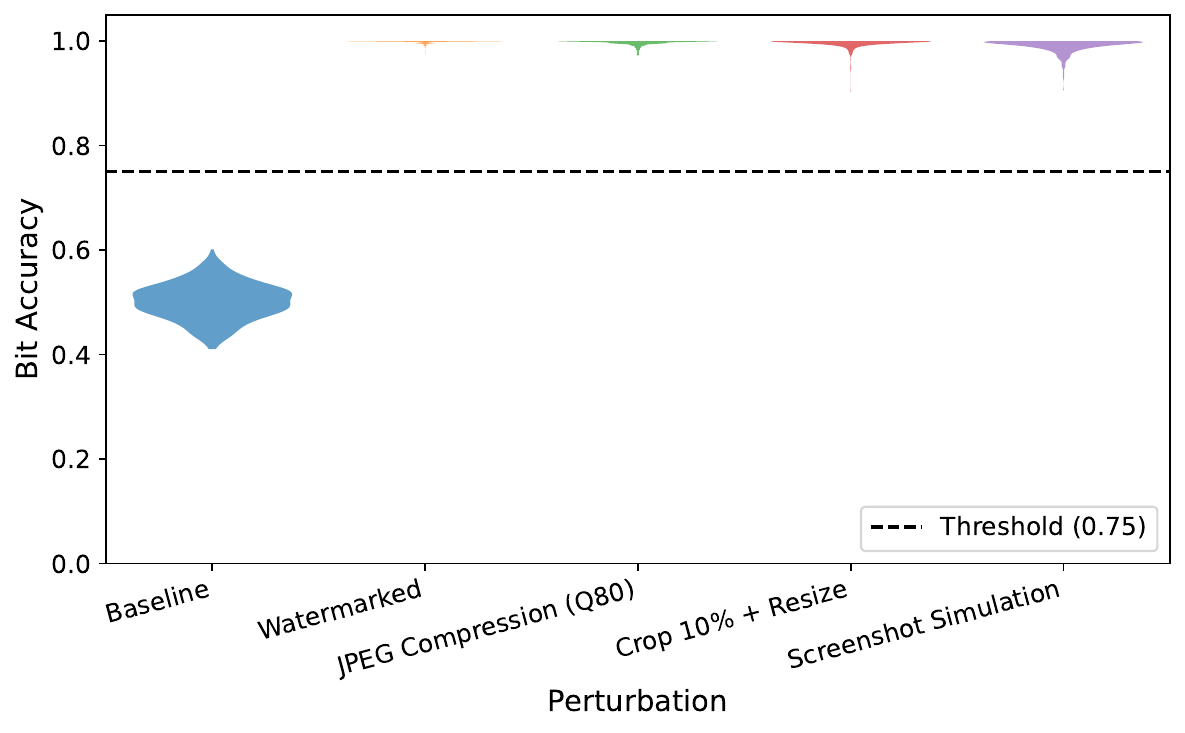}
\caption{Bit accuracy distributions across experimental conditions ($N$$=$500 per condition). The dashed line marks the 0.75 detection threshold. Baseline images (no watermark) cluster at chance level ($\sim$0.50), confirming no false detections. All watermarked conditions remain well above the threshold, with screenshot simulation producing the widest spread (min$=$0.906). No image crosses the detection boundary under any perturbation.}
\label{fig:bit-accuracy}
\end{figure}

The audit protocol in Section~\ref{sec:audit} jointly evaluates C2PA manifest assertions and watermark detection status for each image and classifies it into one of the four conflict-matrix states. A necessary condition for this joint evaluation is that the watermark signal persists through the perturbations applied before signing. Figure~\ref{fig:bit-accuracy} shows bit accuracy distributions across all experimental conditions. JPEG compression at quality 80 produces negligible degradation (mean 0.998, minimum 0.973), while crop-and-resize and screenshot simulation introduce progressively wider spreads with minimum bit accuracies of 0.902 and 0.906, respectively. In all three cases, every image remains above the 0.75 detection threshold, confirming that the \textit{Integrity Clash} survives the realistic editorial and platform-mediated transformations tested in this work.

Table~\ref{tab:audit} reports classification accuracy across all conflict-matrix states. The protocol correctly classifies every image in every condition, including all 2,000 Q4b instances spanning the unperturbed and three perturbed variants. No image from the non-adversarial pipelines is misclassified as an authenticated fake, and no authenticated fake evades detection. The protocol requires no infrastructure beyond what already exists in isolation, a C2PA manifest verifier and a watermark detector. Its contribution is the joint evaluation logic that cross-references the two outputs and flags semantic inconsistencies that neither layer surfaces independently. We note that the perfect classification reported here reflects the strong robustness of Pixel Seal under the perturbations tested; the implications of weaker watermarking schemes are discussed in Section~\ref{limits}.

\begin{table}[t]
\centering
\small
\begin{tabular}{lc}
\toprule
State / Condition & Correctly Classified (\%) \\
\midrule
Q1 (Silent Zone)         & 100.0 \\
Q2 (Fragile Provenance)  & 100.0 \\
Q4a (Verified Synthetic) & 100.0 \\
\midrule
Q4b (Authenticated Fake) & \\
\quad No perturbation    & 100.0 \\
\quad JPEG Q80           & 100.0 \\
\quad Crop 10\% + resize & 100.0 \\
\quad Screenshot sim     & 100.0 \\
\bottomrule
\end{tabular}
\caption{Cross-layer audit protocol evaluation. Each conflict-matrix state contains 500 images; Q4b includes 500 images per perturbation condition (2{,}000 total). Q4b (Authenticated Fake) is the positive class, with TPR$=$1.000, FPR$=$0.000, and Accuracy$=$1.000.}
\label{tab:audit}
\end{table}

\section{Discussion}
\label{sec:discussion}

\subsection{Implications for Deployed Verification Infrastructure}
The results in Section~\ref{sec:evaluation} confirm that the \textit{Integrity Clash} is not an artifact of a contrived experimental setup but a reproducible condition in currently deployed verification workflows. Figure~\ref{fig:verify-screenshots} demonstrates this directly as the Content Credentials Verify tool displays the same AI-generated, watermarked image as either AI-generated or human-edited depending entirely on the attached manifest, with no mechanism to inspect or cross-reference the pixel-level watermark signal. The verifier's sole security concern is certificate trust, not semantic consistency between the metadata and the underlying pixel data.

This behavior is not a flaw in any single implementation but a consequence of how the two verification layers were designed and deployed independently. Each layer operates correctly within its own verification scope. However, no deployed workflow, to our knowledge, jointly evaluates both outputs for consistency. The cross-layer audit protocol evaluated in Section~\ref{sec:CLAP} demonstrates that this joint evaluation is technically straightforward, requiring only that a verifier query both layers and compare their outputs against the consistency rules defined by the conflict matrix. The gap is not technical but organizational where the two verification infrastructures were developed by different communities, standardized through different bodies, and deployed through different integration points, with no specification requiring that one layer condition on the other.

These implications extend directly to workflows that exist today. An AI-generated image carrying an embedded watermark from its generation pipeline~\cite{MetaSeal2026} can be opened in a C2PA-compliant application such as Adobe Photoshop~\cite{adobe2024photoshop_cc}, lightly edited, and exported with a valid Content Credential declaring a human editing action. The current specification does not require the signing application to inspect pixel data for pre-existing watermark signals or to propagate prior origin information into the new manifest. The watermark survives the editorial processing, and the image now carries two competing provenance claims backed by independent verification infrastructures with a cryptographically valid manifest attributing the work to a human editor, and a pixel-level signal identifying it as the output of a generative model. No deployed system adjudicates between them.

The cross-layer audit protocol evaluates only C2PA assertions and watermark status, but its consistency-checking logic is extensible to additional signals. Passive forensic methods such as frequency-domain analysis and learned AI-generation classifiers provide independent evidence of synthetic origin that depends on neither a declared manifest nor an embedded watermark. Integrating these as supplementary inputs would strengthen detection when one primary layer is absent or degraded, and agent-based forensic pipelines~\cite{liang2025evidence} and multi-signal consistency frameworks~\cite{itu2024deepfakes, loth2026origin} offer architectural models for doing so.

\subsection{Limitations \& Future Work}
\label{limits}
As adoption of both systems accelerates, the absence of cross-layer adjudication will become a practical liability for any platform, newsroom, or legal proceeding that relies on either signal in isolation.

Several limitations constrain the scope of these findings. The core experiments use a single watermarking method, Pixel Seal~\cite{souvcek2025pixel}, which is the highest-performing model in the Meta Seal suite~\cite{MetaSeal2026} but may not be representative of all deployed watermarking schemes. Watermarking approaches with weaker robustness properties could fall below the detection threshold under the same perturbations, at which point the misleading manifest becomes the sole remaining signal with no contradicting evidence. This failure mode is arguably more dangerous than the \textit{Integrity Clash} itself, as it is undetectable even by a cross-layer audit. The C2PA manifests were signed with a self-signed research certificate, which caused the Content Credentials Verify tool to flag the issuer as unrecognized. A commercially trusted certificate from a C2PA-compliant application would eliminate this warning while the underlying semantic contradiction persists. Real-world platform testing was partial as we validated against the Content Credentials Verify website but did not test against platform-level moderation pipelines, which may implement additional heuristics beyond manifest validation. Finally, all experiments operate in a controlled setting in which we control both signing and detection, and are restricted to the image modality.

Future work should extend the cross-layer audit to video and audio modalities, where both C2PA and watermarking are being deployed. Equally important is the design of interfaces that communicate cross-layer conflicts to end users rather than surfacing each signal independently. The C2PA specification itself could be extended to require that signing applications inspect data for pre-existing watermark signals before issuing a manifest, closing the semantic gap at the point of signing rather than relying on downstream audit.

\section{Conclusion}
This work formalizes and empirically demonstrates the \textit{Integrity Clash}: a condition in which a cryptographically valid C2PA manifest and a detected AI-origin watermark coexist on the same asset with contradictory provenance claims, passing all deployed verification checks in isolation. The authenticated fakes produced through this workflow require no cryptographic compromise, only the omission of a single assertion field permitted by the current specification. A minimal cross-layer audit protocol that jointly evaluates both signals detects these contradictions with perfect accuracy across all tested conditions, confirming that the gap between the two verification layers is unnecessary and technically straightforward to close. As both systems see wider adoption, any serious content authenticity ecosystem must move from trusting either layer in isolation to auditing the gap between them.

\section*{Acknowledgments}
This work was supported in part by the NSF under CNS-2247795, 2427505, and 2141622, by the Office of Naval Research under N00014-22-1-2680, and by a gift from Cisco.

\newpage

{
    \small
    \bibliographystyle{ieeenat_fullname}
    \bibliography{main}
}

\end{document}